\documentclass[10pt]{article}
\usepackage{amsfonts}
\usepackage{amsmath}
\usepackage{amscd}
\usepackage{amssymb}
\usepackage{graphicx}

\begin{document}
\title{\bf The one dimensional infinite square well with variable mass.}

\author{\bf J.J. Alvarez${}^\dagger$, M. Gadella${}^\ddagger$, L.P. Lara${}^\S$. }

\maketitle

${}^\dagger$E.U. de Inform\'atica de Segovia, 40005, Segovia, Spain,
jjalvarez@infor.uva.es.

${}^\ddagger$Departamento de F\'{\i}sica Te\'orica, At\'omica y
\'Optica, Universidad de Valladolid, 47071 Valladolid, Spain,
manuelgadella1@gmail.com.

${}^\S$Departamento de F\'{\i}sica, FCEIA, Avda. Pellegrini 250,
Rosario, Argentina. Departamento Sistemas, FRRO, Zevallos 1345,
Rosario, Argentina, lplara@fceia.unr.edu.ar.

\begin{abstract}

We introduce a numerical method to obtain approximate eigenvalues
for some problems of Sturm-Liouville type. As an application, we
consider an infinite square well in one dimension in which the mass
is a function of the position. Two situations are studied, one in
which the mass is a differentiable function of the position
depending on a parameter $b$.  In the second one the mass is
constant except for a discontinuity at some point. When the
parameter $b$ goes to infinity, the function of the mass converges
to the situation described in the second case. One shows that the
energy levels vary very slowly with $b$ and that in the limit as $b$
goes to infinity, we recover the energy levels for the second
situation.

\end{abstract}

\section{Introduction}

In this paper, we introduce a method to evaluate eigenvalues for the
some problems of Sturm-Liouville type. As this method should be
valid in order to obtain the energy levels of some Schr\"odinger
equations whose solutions should fulfil given boundary conditions,
we were looking for an explicit example of application. One of the
fields of research in physics which has recently received some
attention is the study of quantum systems with variable mass
\cite{VM}. Situations  combining a one dimensional system with a
mass jump at one point plus a point potential at the same point have
been already studied by our group \cite{AGHN,GHNN,AGN}. However, our
principal aim in these case was looking for solvable or quasi
solvable systems with resonances and the systems under the study did
not have bound states. The simple system having a purely discrete
spectrum capable to bear a non constant mass seems to be the one
dimensional infinite square well. This is the reason why we have
chosen this example in our discussion along the present paper.

As is well know, the Hamiltonian of the one dimensional infinite
square well has the form (with $\hbar=1$)

\begin{equation}\label{1}
H=\frac{-1}{2m}\,\frac{d^2}{dx^2}+V(x)\,,\quad {\rm with}\quad
V(x)=\left\{ \begin{array}{ccc}
               \infty & {\rm if} & x<-c \\[2ex]
               0 & {\rm if} & -c<x<c \\[2ex]
               \infty & {\rm if} & x>c
             \end{array}
  \right.\,.
\end{equation}

In passing, we comment that this Hamiltonian is not self adjoint
unless we fix some boundary conditions at the points $-c$ and $c$
for the wave functions in its domain. As a matter of fact, $H$
defined on its minimal domain has deficiency indices $(2,2)$ so that
its self adjoint extensions are determined by four independent real
parameters. A study of these self adjoint extensions and their
corresponding energy levels is given in \cite{BFV}.

When we introduce a non constant mass a problem arises connected
with the non commutation between the position and the momentum. In
order to avoid it, one should replace the usual kinetic term in the
Hamiltonian $K=p^2/(2m)$ by the following symmetric expression:

\begin{equation}\label{2}
K=-\frac{1}2\,\left\{m^\alpha(x)\,\frac{d}{dx}\,m^\beta(x)\,\frac{d}{dx}\,m^\alpha(x)
 \right\}\,,
\end{equation}
with with $2\alpha+\beta=-1$. Thus, our Hamiltonian should be
$H=K+V(x)$ with $K$ as in (\ref{2}) and $V(x)$ as in (\ref{1}).

However, this situation is too general and we should make one choice
being both simple and natural. The choice that looks more natural is
probably $\alpha=0$, $\beta=-1$. The connection of this choice with
the conservation of Galilei invariance was discussed in \cite{LL}
and on this Galilei invariance is already a basis to make it.

Assume first that the mass is a differentiable function $m(x)$ of
the position. Then after a simple calculation, one shows that the
time independent Schr\"odinger equation $H\varphi=E\varphi$ is given
inside the interval $[-c,c]$ by

\begin{equation}\label{3}
 \frac1{m(x)}\,\varphi''(x)-\frac{m'(x)}{m^2(x)}\, \varphi '(x)+2E\varphi(x)=0
\end{equation}
and zero otherwise. Here, we may use the standard boundary condition
at the borders: $\varphi(-c)=\varphi(c)=0$ to obtain solutions of
the eigenvalue problem.

A second possibility is the assumption of a constant mass except for
a discontinuity at some point, say $a\in[-c,c]$ (we may take $a=0$,
but also there is no need for the interval to be $[-c,c]$, it may
also be $[0,c]$ or any other). In this case, along boundary
conditions at the border, we should give matching conditions at the
point $a$. Here, we use again $\varphi(-c)=\varphi(c)=0$. Then, we
need to fix the matching conditions at $a$ (we henceforth assume
that $a=0$ for simplicity). In \cite{AGHN}, it was shown that the
choice  $\alpha=0,\beta=1$ along the additional condition

\begin{equation}\label{4}
\frac1{m_2}\,\varphi'(0+)-\frac1{m_2}\,\varphi'(0-)=0\,,
\end{equation}
where $\varphi'(0+)$ and $\varphi'(0-)$ are the right and left
limits of the wave function $\varphi(x)$ at the origin, gives the
following expression for the kinetic term $K$:

\begin{equation}\label{5}
K=\left\{ \begin{array}{ccc}
            -\frac1{2m_1} & {\rm if} & x<0 \\[2ex]
           -\frac1{2m_2} & {\rm if} & x>0
          \end{array}
   \right.\,.
\end{equation}

The problem of the self adjoint extensions of (\ref{5}) was
discussed in \cite{GKN}. In order to fix a proper self adjoint
choice for our Hamiltonian, in addition to the boundary values at
the borders we need to settle matching conditions at the point at
which we have the mass discontinuity. These conditions are relations
between the left and right limits of the function and its first
derivative. Then along (\ref{4}), we can use the continuity of the
wave function. The resulting matching conditions can be written as

\begin{equation}\label{6}
\left(
  \begin{array}{c}
    \varphi(0+) \\[2ex]
   \varphi'(0+) \\
  \end{array}
\right)= T  \left(
  \begin{array}{c}
    \varphi(0-) \\[2ex]
   \varphi'(0-) \\
  \end{array}
\right)= \left(
            \begin{array}{cc}
              1 & 0 \\[2ex]
              0 & \frac{m_2}{m_1} \\
            \end{array}
          \right)
   \left(
  \begin{array}{c}
    \varphi(0-) \\[2ex]
   \varphi'(0-) \\
  \end{array}
\right)\,.
\end{equation}

The self adjointness of the Hamiltonian is determined by both,
boundary and matching conditions. In particular, matching conditions
are determined by the matrix $T$ in (\ref{6}), which should verify
the following relation \cite{AGHN,GHNN}:

\begin{equation}\label{7}
M_1=T^\dagger M_2T\,,\quad M_i=\frac1{2m_i}\;\left(
  \begin{array}{cc}
0 & 1 \\[2ex]
-1 & 0 \\
   \end{array}
     \right)\,,\quad i=1,2\,.
\end{equation}
$T^\dagger$ denotes the adjoint matrix of $T$.

The general solution of the Schr\"odiger equation under the above
conditions is given by

\begin{equation}\label{8}
\varphi(x)=A\sin(k_1(c+x))H(-x)+B\sin (k_2(c-x))H(x)\,,
\end{equation}
where $A$ and $B$ are constants, $H(x)$ is the Heaviside step
function and

\begin{equation}\label{9}
 k_i=\sqrt{2m_iE}\,,\quad i=1,2\,.
\end{equation}
Note that $\varphi(-c)=\varphi(c)=0$ so that the required  boundary
conditions are satisfied. The given (\ref{6}) matching conditions
applied to (\ref{8}) give:

\begin{eqnarray}
A\sin(k_1c)=B\sin(k_2c) \Longrightarrow B=A\,\frac{\sin(k_1c)}{\sin(k_2c)} \label{10}
\\[2ex]
\frac{Ak_1}{m_1}\,\cos(k_1c)=\frac{Bk_1}{m_2}\,\cos(k_2c)
\Longrightarrow
\frac{k_1}{m_1}\,\cos(k_1c)=\frac{k_1}{m_2}\,\frac{\sin(k_1c)}{\sin(k_2c)}
\,\cos(k_2c)\,.\label{11}
\end{eqnarray}
This and (\ref{9}) give the following transcendental equation for
$E$:

\begin{equation}\label{12}
\sqrt{m_2}\;\tan[c\sqrt{2m_2E}]=\sqrt{m_1}\;\tan[c\sqrt{2m_1E}]\,.
\end{equation}

It is noteworthy that equation (\ref{12}) gives the energy levels
for any one dimensional infinite square well of width $2c$ with a
jump of mass at the middle point and masses $m_1$ and $m_2$ (with
jump $\Delta:=m_2-m_1$), because of this system is invariant under
translation. Therefore,  formula (\ref{12}) given the energy levels
is valid if the interval were $[0,2c]$ instead $-c,c]$, for
instance. We shall take into account this fact in our explicit
calculations.

This paper is organized as follows: In Section 2, we introduce the
method for obtaining the energy levels and in Section 3, we compare
our method to other methods. In Section 4, we apply our method to a
situation in which the dependence of the mass with the position is
differentiable. Furthermore, the function of mass depends on an
additional parameter $b$ such that in the limit $b\longmapsto\infty$
we obtain the coordinate dependence of the mass given by
$m(x)=m_1H(-a)+m_2H(x)$, i.e., a constant mass except for a
discontinuity at one point, say $a$. This is interesting as the
energy levels for the continuous distribution of mass converge to
the energy levels for the limit case.

\section{A method for the determination of the eigenvalues.}

Along the present section, we shall develop a method to solve
numerically the equations that will arise in the problem described
here. These equations are of the of the following type:

\begin{equation}\label{13}
y''(x)=F_{\lambda}\,(x,\,y(x),\,y'(x)\,)\,.
\end{equation}
This equation depends explicitly of the eigenvalue $\lambda$ to be
determined, as is exactly the case of the (time independent)
Schr\"odinger equation. Here, $x\in I\equiv [a,b]$. We assume that
the following conditions hold:

\begin{equation}\label{14}
c_{1}\,y(a)+c_{2}\,y'(a)    =c_{3}\,, \qquad
d_{1}\,y(b)+d_{2}\,y'(b) =d_{3}\,,
\end{equation}
where the constants $c_k$ and $d_k$, $k=1,2,3$ are known. Note that
this problem is just a generalization of Sturm-Liouville.

A variety of methods for solving this problem ((\ref{13}) with
(\ref{14})) exist. One of the most efficient is the Differential
Transformation Method (DTM)\cite{DTM,DTM1,DTM2,DTM3,DTM4}, which is
clearly described in \cite{DTM4}. This method is beautiful, elegant
and, in many cases, effective. However,  its practical
implementation requires numerical determination of roots, a
calculation which is sometimes tedious, particularly in the case of
non-linear equations.

We are introducing another method to solve numerically ordinary
differential equations, which has been suggested to us by our
experience in the use of the software Mathematica. A version of this
method was designed to approximate periodic solutions \cite{GL}.

We hereby describe the method and use it for the numerical study of
the infinite square well with a variable mass.

Our method is based in the approximation of the solution $y(x)$ by
means of Taylor polynomials whose order depends on the desired
degree of accuracy. In order to perform the expansion on the finite
Taylor series, we need to impose the condition that
$F_\lambda(x,y,y')$, where $\lambda$ denotes a parameter, be
analytic or at least differentiable up to the necessary order on the
interval $[a,b]$. Contrary to the usual procedure in the DTM,  we do
not determine algebraically the coefficients in the Taylor
polynomial, but instead we calculate them analytically, often with
the help of a software for symbolic calculus.

The second difference between our method and the DTM can be
explained as follows: when searching for solutions to ODE with given
boundary conditions we often find periodic solutions which cannot be
accurately approximated by a unique Taylor polynomial on the whole
integration interval. When solving (\ref{13}) subject to conditions
(\ref{14}), this problem has the following cure:

Let us divide the integration interval $[a,b]$ into $m$ subintervals
with the same length by taking the points $x_k=a+kh$ with
$h=(b-a)/n$, $k=1,2,\dots,m$. Here the number $m$ of subintervals
depends on the precision we want. Then, let us write (\ref{13}) as
the following system:

\begin{equation}\label{15}
y'(x)=z(x)\,,\qquad z'(x)=F_\lambda(x,y(x),y'(x))\,.
\end{equation}
Then on each interval $(x_k,x_{k+1})$, we approximate the functions
$y(x)$ and $z(x)$ by respective Taylor polynomials as follows:

\begin{equation}\label{16}
 y_n(x)=\sum_{j=0}^n \frac1{j!}\,y^{(j)}(x_k)(x-x_k)^j\,,
 \quad  z_n(x)=\sum_{j=0}^n \frac1{j!}\,z^{(j)}(x_k)(x-x_k)^j\,,
\end{equation}
where we determine the derivatives $y^{(j)}$ and $z^{(j)}$ by means
of (\ref{15}), i.e., $y^{(1)}(x)=z(x)$,
$y^{(2)}(x)=F_\lambda(x,y(x),z(x))$, $y^{(3)}(x)=\partial
F_\lambda/\partial x+z\partial F_\lambda/\partial
y+F_\lambda(\partial F_\lambda/\partial z)$ and so on up to the
desired derivative. Similarly, we obtain the successive derivatives
of $z(x)$.

Then, starting with the first equation in (\ref{15}), we obtain the
initial values to obtain $y_n(x)$ and $z_n(x)$ on the interval
$[x_0,x_1]$, which are parameterized by the eigenvalue $\lambda$.
Then, we know the values $y_n(x_1)$ and $z_n(x_1)$ which allow us to
construct $y_n(x)$ and $z_n(x)$ on the interval $[x_1,x_2]$ and so
on.

After obtaining the approximation for the last interval, we note
that the second equation in (\ref{13}) gives us the condition
$d_1y(x_m)+d_2z(x_m)=d_3$ with $x_m=b$, which gives an algebraic
equation in $\lambda$. Then, the roots of this equation are the
eigenvalues we search for.

Once we have chosen one of these eigenvalues, the segmentary
solutions are well defined. The values of $n$ and $m$ should be
determined by practice depending on the desired accuracy.

For the above calculations we use the software Mathematica.

\section{Comparison with results obtained by other authors.}

Along the present section, we compare our results to the results
obtained by the DTM method as introduced in \cite{DTM,DTM1}. Take
for instance a problem considered by Chao Kuang Chen and Shing Huei
Ho in \cite{DTM}, the differential equation with boundary conditions
given by

\begin{equation}\label{17}
y''(x)+\lambda x^2y(x)=0\,, \quad y(0)=y(1)=0\,.
\end{equation}
Here, the solution is given by

\begin{equation}\label{18}
y(x)=\frac{\mathit{\Gamma}(1/4)}{2\times 2^{1/4}\lambda^{1/4}
\sqrt{\pi}}\;(\mathit{D}_{-1/2}\,(\,(-1+i)\,\lambda
^{1\,/\,4}\,x\,)-\mathit{D}_{-1/2}\,(\,(1+i)\,\lambda^{1\,/\,4}\,x\,)\,)\,,
\end{equation}
where $D_\nu(x)$ is the parabolic cylindric function
\cite{AS} of index $\nu$ and $\Gamma(x)$ is the Gamma function.
Starting with (\ref{18}) and the roots of the transcendental
equation $y(0)=1$, we obtain numerical values for the eigenvalues
$\lambda_{exact}$. Note that Taylor expansion of (\ref{18}) on a
neighborhood of zero has only the odd powers $x^{4k-3}$ with
$k=1,2\dots$.

Now, let us go back to the method introduced in the previous
section. Take the interval $[0,1]$ and divide it into only one
interval, so that $m=0$. Then, let us determine the solutions of
(\ref{17}) for different values of $n$. First of all, let us
determine the the smallest eigenvalue. Its ``exact'' value is
30.9333. Then, we obtain

\bigskip

\[
\begin{array}
[c]{cccccc}
n & 13 & 17 & 21 & 25 & 29\\[2ex]
\lambda_{n} & 30.2370 & 30.9921 & 30.9303 & 30.9335 & 30.9333\\[2ex]
\Delta y & - & 0.75 & -0.062 & 0.0032 & -0.0002\\[2ex]
\epsilon_{r}\% & 2 & 2\,10^{-1} & 1\,10^{-2} & 5\,10^{-4} &
1\,10^{-4}
\end{array}
\]

\medskip
\centerline{Table 1}

\bigskip

Here, $\Delta y$ denotes the differences
$\lambda_{17}-\lambda_{13}$, then, $\lambda_{21}-\lambda_{17}$ and
so on and $\epsilon_{r}\%$ is the modulus of the relative difference
between $\lambda_{exact}$ and the eigenvalue $\lambda_n$ as given in
the above table. These results are equivalent to those obtained in
\cite{DTM}. In order to implement our method, we have used
Mathematica and the CPU times in a regular computer are less than a
second. Since the exact solution is known, we can estimate the
difference between solutions by means of the error parameter, here
named as error for simplicity, defined as

\begin{equation}\label{19}
D_n=\int_0^1 (y_{exact}(x)-y_n(x))^2\,dx\,.
\end{equation}

For instance, for the first eigenvalue, according to Table 1, we
obtain: $D_{13}=2\,10^{-5}$ and $D_{12}=4\, 10^{-10}$. We see that
this error is quite small.

Using (\ref{16}), we can obtain the approximate eigenfunction for
$n=21$, which is

\begin{equation}\label{20}
y=x-1.54652\, x^{5}\,+0.664364\, x^{9}\,-0.131724\,
x^{13}+0.0149789\, x^{17}-0.0011031\, x^{21}\,,
\end{equation}
while the result obtained in \cite{DTM} was:

\begin{equation}\label{21}
y=x-1.5465\,x^{5}+0.664351\,x^{9}-0.13172\,x^{13} +0.0149783\,
x^{17}-0.00110305\,x^{21}\,.
\end{equation}
Here, the authors obtain $D_{21}=5\,10^{-10}$ a result that our
method slightly improves.

For the second eigenvalue, whose ``exact'' result is 139.530, we
obtain:

\bigskip

\[
\begin{array}
[c]{cccccc}
n & 29 & 33 & 37 & 41 & 45\\[2ex]
\lambda_{n} & 141.889 & 139.315 & 139.549 & 139.529 & 139.530\\[2ex]
\Delta y & - & -2.6 & 0.21 & -0.02 & 0.001\\[2ex]
\epsilon_{r}\% & 2\,10^{-2} & 1\,10^{-3} & 1\,10^{-4} & 8\,10^{-6} &
1\,10^{-6}
\end{array}
\]

\medskip
\centerline{Table 2}
\bigskip

Here, the errors are given by $D_{29}=1\,10^{-5}$ and
$D_{45}=2\,10^{-14}$.

Now, we want to estimate the third eigenvalue. In the DTM, we need
to increase the degree $n$ of the polynomial. With our method, we
can make use of the numbers $m$ and $n$ to obtain a sufficient
accuracy for a bigger number of eigenvalues. For instance, if we
take $m=10$ and $n=10$, we can obtain the values of the seven first
eigenvalues of a reasonable precision. In the next table, we give
the errors that our method shows in relation with the exact results
for the seven first eigenvalues:

\bigskip

\[
\begin{array}
[c]{cccccccc} eigenvalue & 1 & 2 & 3 & 4 & 5 & 6 & 7 \\[2ex]
\epsilon_{r}\% & 2\,10^{-4} & 7\,10^{-5} & 2\,10^{-4} & 3\,10^{-3} &
2\,10^{-2} & 1\,10^{-2} & 6\,10^{-1}\,.
\end{array}
\]

\medskip
\centerline{Table 3}
\bigskip

Finally, we shall compare our method to the standard method of the
determination of the eigenvalues by the method of the polynomial
\cite{}. This is a numerical resolution of equation (\ref{15}) by
replacing the second derivative by an approximation by finite
divided and centered differences. This generates a homogenous system
of equations. The expansion of the determinant of the coefficient
matrix produces a polynomial such that its real roots are the
desired eigenvalues.

For the first eigenvalue, the method of the polynomial gives the
following results:

\[
\begin{array}
[c]{cccc}
n & 200 & 500 & 1000\\[2ex]
\lambda & 30.9324 & 30.9332 & 30.9333\\[2ex]
\epsilon_{r}\% & 3\,10^{-3} & 5\,10^{-4} & 1\,10^{-4}\,.
\end{array}
\]

\medskip
\centerline{Table 4}
\bigskip

In Table 4, $n$ is the subinterval given in the divided difference.
The corresponding results for the second eigenvalue are:

\bigskip

\[
\begin{array}
[c]{cccc}
n & 200 & 500 & 1000\\[2ex]
\lambda & 139.510 & 139.527 & 139.529\\[2ex]
\epsilon_{r}\% & 1\,10^{-2} & 3\,10^{-2} & 5\,10^{-4}\,.
\end{array}
\]

\medskip
\centerline{Table 5}
\bigskip

From the results of Table 4 and Table 5, we conclude that in order
to obtain the level of precision reached in the results given in
Table 1 and Table 2, the polynomial method requires a large number
of nodes than in the method proposed in Section2. In addition, CPU
times are significatively higher. Furthermore, if we want to find
the eigenvalues next to the second, we need to add more nodes and
consequently the CPU time.

As a conclusion, we have introduce an alternative to calculation to
powerful methods such that, by using the software of symbolic
calculus (we have used Mathematica), it shows a really simple
implementation that can be even useful for a beginner. In addition,
it has the advantage that the few former eigenvalues can be obtained
with a Taylor polynomial of rather low degree and small CPU times.

\section{Infinite square well with variable mass.}

Along this section, we shall apply the numerical formalism here
introduced in the study of the energy levels of the infinite square
well with a variable mass. We consider two possibilities:

i.) A mass jump as described in the Introduction. If the mass jump
is located at the point $a$, the mass function in terms of the
position can be written, inside the interval with zero potential as
$m_\infty(x)=m_1H(a-x)+m_2H(x-a)=m_1+(m_2-m_1)H(x-a)$, where
$H(\omega)$ is the Heaviside step function.

ii.) As a previous exercise, we shall approximate the mass jump by a
differentiable mass function $m_b(x)$, depending on a parameter $b$,
with the condition

\begin{equation}\label{22}
\lim_{b\mapsto\infty}m_b(x)=m_\infty(x)\,.
\end{equation}

Thus, let us study first the Schr\"odinger equation (\ref{3}) in
which the well interval has been chosen to be $[0,1]$ instead of
$[-1/2,1/2]\,, c=1$ for simplicity in our calculations. Note that
the energy levels should be invariant under space translations, so
that our results will be equally valid for any interval of width 1.

Now, we construct the function of mass $m_b(x)$ as follows: Take the
function

\begin{equation}\label{23}
g_b(x):=\frac1{1+e^{-b(x-1/2)}}\,.
\end{equation}
Observe that

\begin{equation}\label{24}
\begin{array}{ccc}
{\rm if} & x>1/2 & \lim_{b\mapsto\infty}g_b(x)\longmapsto 1
  \\[2ex]
{\rm if} & x=1/2& g_b(1/2)=1/2 \\[2ex]
{\rm if} & x<1/2 & \lim_{b\mapsto\infty}g_b(x)\longmapsto 0\,.
\end{array}
\end{equation}
Therefore,

\begin{equation}\label{25}
\lim_{b\mapsto\infty}\frac1{1+e^{-b(x-1/2)}}=H(x-1/2)
\end{equation}
pointwise. Consequently, if we define

\begin{equation}\label{26}
m_b(x):=m_1+\frac{m_2-m_1}{1+e^{-b(x-1/2)}}
\end{equation}
then,

\begin{equation}\label{27}
\lim_{b\mapsto\infty}m_b(x)=m_1+(m_2-m_1)H(x-1/2)=m_\infty(x)\,.
\end{equation}

Now, take $m_b(x)$ as defined in (\ref{26}) and use it as a
particular realization of $m(x)$ in the Scr\"odinger equation
(\ref{3}). We intend to solve numerically this equation for given
values of the parameters. Our first choice is $m_1=1$,
$\Delta:=m_2-m_1=1$ and $b=10$. We have to use the boundary values
of wave functions at the points $0$ and 1. To be consistent with the
comments in the Introduction, these boundary values should be
$\varphi(0)=\varphi(1)=0$. Then, we determine the solutions using
the method described in Sections 2 and 3.

Then, in order to initiate the iterative process as described in
equation (\ref{16}), let us take the initial values
$y(0)=\varphi(0)=0$ and $z(0)=\varphi'(0)=1$ (observe that the value
of the derivative $\varphi'(0)$ can be somehow arbitrarily chosen as
it affects to the norm of the wave function only). Following Section
2,  results depend on two parameters: $n$ the chosen degree of the
polynomial and $m$ the number of intervals which divide the
integration domain (here $[0,1]$).

In the next table, we show as an example the values of the first
$E_1$ and the fifth $E_5$ eigenvalues of the energy in terms of the
parameters $n$ and $m$. The first column gives the values of $n$ and
the first row the values of $m$.

Thus, for $E_1$, we have

\bigskip

$\qquad
\begin{array}
[c]{ccccccc}
n\setminus m & 50 & 60 & 80 & 100 & 300 & 600\\[2ex]
1 & 3.5540 & 3.5417 & 3.52712 & 3.51878 & 3.49804 & 3.49321\\
2 & 3.48446 & 3.48567 & 3.48690 & 3.48745 & 3.48837 & 3.48845\\
3 & 3.48841 & 3.48844 & 3.48847 & 3.48847 & 3.48848 & 3.48848\,.
\end{array}
$

\medskip
\centerline{Table 6}
\bigskip

For $E_5$ the values we have obtained are

\bigskip

$\qquad
\begin{array}
[c]{ccccccc}
n\setminus m & 50 & 60 & 80 & 100 & 300 & 600\\[2ex]
1 & 91.6682 & 89.3794 & 87.1101 & 86.0380 & 84.1501 & 83.8877\\
2 & 81.1658 & 81.8829 & 82.6441 & 83.0145 & 83.6271 & 83.6867\\
3 & 83.6004 & 83.6506 & 83.6860 & 83.6970 & 83.7064 & 83.7066\,.
\end{array}
$

\medskip
\centerline{Table 7}
\bigskip

In both tables, we can appreciate the convergence either if we fix
$n$ and increase $m$ or viceversa. As one could have expected, the
bigger $n$ the smaller the number of $m$ to achieve similar
accuracy.

If we take the resulting values for $n=3$ and $m=600$ as reference
values, then we may compare the percentual relative difference
between the reference value and those obtained for $m=50$. The
results are given in the following table:

\bigskip

\begin{center}

$
\begin{array}
[c]{ccc}
n & E_{1} & E_{5}\\[2ex]
1 & 2\,\% & 10\,\%\\
2 & 0.1\,\% & 3\,\%\\
3 & 0.001\,\% & 0.1\,\%\,.
\end{array}
$

\end{center}

\medskip
\centerline{Table 8}
\bigskip

This shows that for the five first eigenvalues of the energy, at
least, a good choice could be $n=3$, $m=50$. It is also interesting
to remark that the approximation given by formula (\ref{28}) below
gives $6\%$ for $E_1$ and $2\%$ for $E_5$. This choice gives CPU
times lower than two seconds in a computer AMD Athlon II X2 250 3.00
GHz, RAM 4 GB using Mathematica software.

\begin{figure}
\begin{center}
\includegraphics[width=0.6\textwidth]{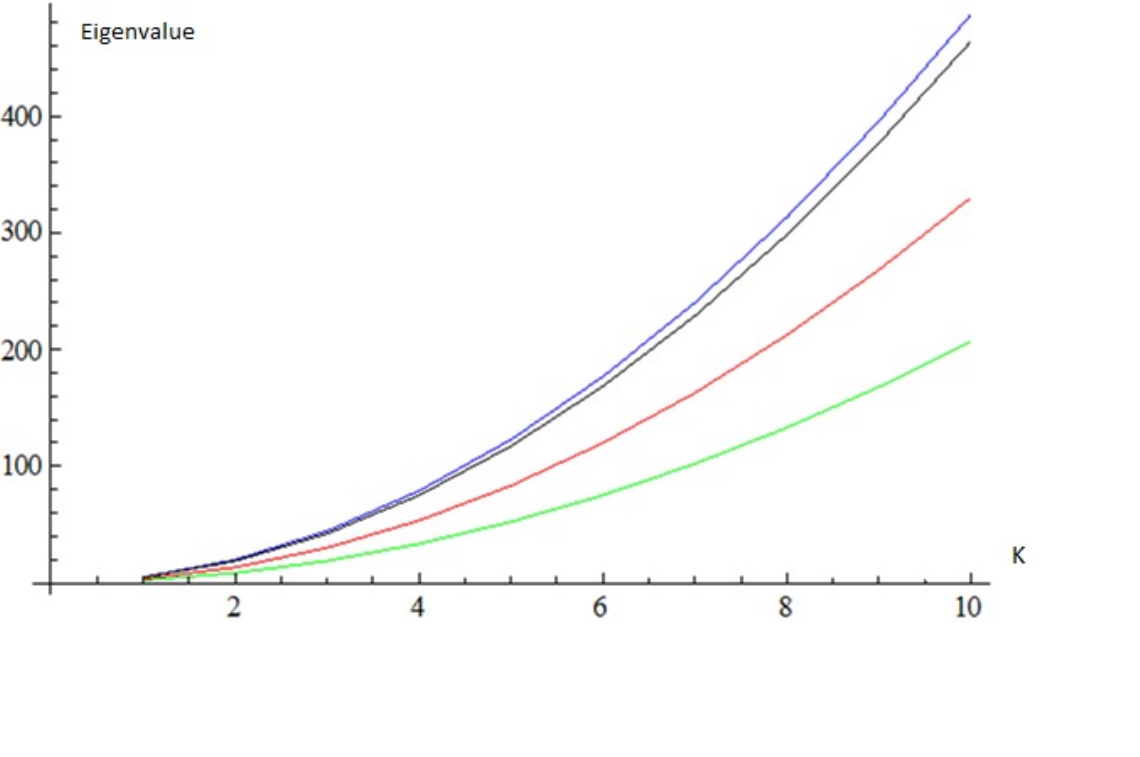}\quad
\caption{Energy levels for different values of $\Delta$.   \label{f1}}
\end{center}
\end{figure}

\subsection{Dependence of the mass on the parameter $b$.}

Next, we analyze the behavior of the energy spectrum under changes
on the values of the mass parameters. Take again $m_1=1$ and let us
evaluate the ten first energy levels for different values on the
difference $\Delta=m_2-m_1$. We have moved $b$ along the interval
$(0.05,200)$ and $\Delta$ along $(0.0,3.0)$. The dependence of the
energy levels does not depend significantly on $b$ within the
considered interval. Contrarily, the variation of the energy  levels
with $\Delta$ behave according to the law

\begin{equation}\label{28}
E_k\approxeq \frac{(k\pi)^2}{2+\Delta}\,,\quad k=1,2,\dots,10\,,
\end{equation}
with a relative error smaller than $8\%$.

In Figure 1, we show the value of the energy levels for different
values of $\Delta$. The blue curve corresponds to $\Delta=0$
(constant mass). Below, we plot the curves for $\Delta=0.1\,, 1.0$
and $3.0$. Curves obtained with formula (\ref{28}) match with curves
obtained numerically. This Figure remains essentially unaltered if
we modify the values of $b$ from $0.05$ to $50$. Although the energy
spectrum is discrete this continuous representation of the energy
levels seem to be very explicative by itself.

\subsection{Constant mass with jump at $x=1/2$.}

We discuss here the model of the infinite square well with a mass
dependence on the position given by $m_\infty(x)$ as in (\ref{27}).
The equation we have to solve here is (\ref{12}) with $c=1/2$. Let
us choose $m_1=1$. Then, the particular form of (\ref{12}) becomes:

\begin{equation}\label{29}
\sqrt{m_2}\;\tan\left[ \sqrt{m_2} \;\sqrt{\frac E2}
\right]=\tan\left[\sqrt{\frac E2}  \right]
\end{equation}

In the simple case in which we choose $m_2=n^2$, i.e., the square of
a natural number, this equation is explicitly solvable and gives:

\begin{equation}\label{30}
E_k=2(k\pi)^2\,,\quad k=0,1,2,\dots\,.
\end{equation}

In addition, if we choose the mass jump $\Delta=m_2-m_1$ to be
smaller than three, the energy levels can be approximated quite
reasonably by the following formula:

\begin{equation}\label{31}
E_{k}^*=\frac{(k\pi)^{2}}{2+\Delta}\,, \quad  k=0,1,2,\dots\,.
\end{equation}

Take, for instance $\Delta=1$ ($m_2=2$) and compare the roots of
(\ref{31}) to the roots obtained from (\ref{29}). We compare these
results on Table 9, in which $E_r\%$ determines the percentual
relative variation of $E_k^*$ with respect to $E_k$:

\bigskip

$
\begin{array}
[c]{ccccccccc}
k & 1 & 2 & 3 & 4 & 5 & 10 & 15 & 20\\[2ex]
E_{k} & 3.5846 & 12.909 & 31.602 & 52.937 & 85.448 & 341.514 &
328.987 &
1362.00\\[1ex]
E_{k}^* & 3.290 & 13.159 & 29.609 & 52.638 & 82.247 & 328.99 &
740.22 &
1316.00\\[1ex]
E_{r}\% & 8 & 2 & 6 & 1 & 4 & 4 & 4 & 3
\end{array}
$

\medskip
\centerline{Table 9}
\bigskip

Another example is given in Table 10. Take the first eigenvalue of equation
(\ref{29}) as given by (\ref{31}). This eigenvalue is $E_1^* = 3.5846$. Let us use the method
described in Section 2 for the determination of the eigenvalue of (\ref{3}) with mass
dependence in the coordinate given by (\ref{26}) and $\Delta=m_2-m_1=1$. We use
the integration parameters $n = 3$ and $m = 600$. On Table 10, we show the
dependence of the value of the  first energy level $E_1$ in terms of the parameter $b$
in (\ref{26}). We also include the relative error, $\varepsilon_r\%$ defined above.

The results are:

\bigskip

$
\begin{array}
[c]{cccccc}
b & 10 & 50 & 100 & 200 & 500 \\[2ex]
E_{1} & 3.48847 & 3.57912 & 3.5835 & 3.58423 & 3.58460 \\[1ex]
\varepsilon_{r}\% & 3 & 2.\,10^{-1} & 3.\,10^{-2} & 1.\,10^{-2} & 0 \end{array}
$

\medskip
\centerline{Table 10}
\bigskip

We note how close are the results obtained by our method to the empirical
results given by formulas (\ref{29}) and (\ref{31}).

\section{Concluding remarks}

We have proposed one method to obtain the energy levels of a given Hamiltonian
with purely discrete spectrum. We have applied it to calculate the energy levels
of the one dimensional infinite square well with variable mass. The mass is
written as a function of position. We have considered two possibilities, one in
which the mass is a continuous and even differentiable function of the position
and in the other the mass is constant except for a jump at the middle of the
well. In addition, the continuous function of mass depends on one parameter
so that when this parameter goes to infinite the function of mass goes to the
second case of constant mass with a jump. Our numerical calculations show
that also the energy levels for the continuous mass case go to the energy levels
for the mass jump case.

\section*{Acknowledgements}

Partial financial support is acknowledged to the Spanish Ministry of
Science and Technology (Project MTM2009-10751) and  the Project
UNR-ING 195 (Argentina).

\end{document}